# Decomposition of Amino Acids in Water with Application to In-Situ Measurements of Enceladus, Europa and Other Hydrothermally Active Icy Ocean Worlds

Ngoc Truong[1*], Adam A. Monroe[2], Christopher R. Glein[3], Ariel D. Anbar[2,4], Jonathan I. Lunine[5].

[1]Department of Earth and Atmospheric Sciences, Cornell University, Ithaca, NY 14850
[2]School of Earth and Space Exploration, Arizona State University, Tempe, AZ 85287
[3]Space Science and Engineering Division, Southwest Research Institute, San Antonio, TX 78228
[4]School of Molecular Sciences, Arizona State University, Tempe, AZ 85281
[5]Department of Astronomy, Cornell University, Ithaca, NY 14850

**\*** Corresponding author. Email: ntruong@astro.cornell.edu

**Abstract.** To test the potential of using amino acid abundances as a biosignature at icy ocean worlds, we investigate whether primordial amino acids (accreted or formed by early aqueous processes) could persist until the present time. By examining the decomposition kinetics of amino acids in aqueous solution based on existing laboratory rate data, we find that all fourteen proteinogenic amino acids considered in this study decompose to a very large extent (>99.9%) over relatively short lengths of time in hydrothermally active oceans. Therefore, as a rule of thumb, we suggest that if amino acids are detected at Enceladus, Europa, or other hydrothermally active ocean worlds above a concentration of 1 nM, they should have been formed recently and not be relicts of early processes. In particular, the detection of aspartic acid (Asp) and threonine (Thr) would strongly suggest active production within the ocean, as these amino acids cannot persist beyond 1 billion years even at the freezing point temperature of 273K. Identifying amino acids from the oceans of icy worlds can provide key insight into their history of organic chemistry.

# 1. INTRODUCTION

Saturn's moon Enceladus harbors a global liquid water ocean beneath its icy crust (Thomas et al., 2016) which erupts plumes of gases and grains (both icy and refractory) into space above the south polar region (Porco et al., 2014). The detection of nanometer-sized silica-rich grains from Enceladus suggests ongoing hydrothermal activities in the ocean (Hsu et al., 2015). It has been argued that the only plausible mechanism for forming these grains is via high-temperature water-rock interactions, where ocean water is heated to at least 90°C in the presence of silicate minerals by fluid circulation in the core of Enceladus (Hsu et al., 2015). In the presence of reduced iron, this process can produce molecular hydrogen by serpentinization of rocks, as observed in some hydrothermal systems on Earth (e.g., Lost City, Kelley et al., 2001), and $H_2$ was detected in plumes of Enceladus by the INMS (Ion Neutral Mass Spectrometer) onboard the Cassini spacecraft (Waite et al., 2017). The plumes also contain organic materials (Waite et al., 2009, Postberg et al., 2018) including macromolecular organic compounds of greater than 200 mass units that contain unsaturated carbon atoms with O- or N-bearing groups (Postberg et al., 2018). These lines of evidence make Enceladus a compelling place to search for life in the Solar System (Lunine, 2017).

The Jovian moon Europa is another ocean world in the Solar System. A global salty ocean was found indirectly by the magnetometer instrument onboard the Galileo spacecraft (Khurana et al., 1998, Kivelson et al., 2000). Recently, in-situ evidence of a plume on Europa was discovered by reanalyzing the magnetic field and plasma wave observations of the Galileo spacecraft from the closest encounter flyby of the moon (Jia et al., 2018). If plumes are present (Roth et al., 2014, Sparks et al., 2017) and can be confirmed to come from the subsurface ocean, then many details of the ocean can be revealed through plume measurements, similar to what was done with Cassini for Enceladus (Waite et al., 2009, Postberg et al., 2011). However, details of the physical and chemical properties of the ocean, including the possibility of hydrothermal activity (e.g. Lowell & DuBose, 2005, Vance et al., 2007, Vance et al., 2016) and the habitability of Europa's ocean (Hand et al., 2009) must await further exploration to be conducted by NASA's planned Europa Clipper mission as well as ESA's planned JUICE mission.

On Earth, the alkaline, serpentinizing hydrothermal systems which today host ancient metabolisms like acetogenesis and methanogenesis (e.g., Lost City, Kelley et al., 2001) have been proposed as candidate sites for the origin of life (e.g. Baross & Hoffman, 1985, Russell & Hall, 2006, Martin et al., 2008, Russell et al., 2014). Prebiotic reactions on early Earth could occur also in these same type of vent environments (Russell et al., 2004) and the synthesis and polymerization of amino acids may represent a key step in that process (Huber and Wächtershäuser, 1998, Huber et al., 2003, Lemke et al., 2009). The possible extraterrestrial ocean worlds vents may be similar to these serpentine-hosted vents on Earth, and make icy ocean worlds valuable sites to investigate for the presence of life or prebiotic precursors that may lead to life.

Because proteins catalyze the chemistry of life as we know it, they are indispensable components of the cell and their amino acid building blocks are often considered to be one of the most important biosignatures in searching for in situ evidence of life on other worlds (Neveu et al., 2018). However, before a biological origin can be deduced from any future findings of amino acids at ocean worlds, other non-biological sources must be considered including 1) accretion of primordial materials, and 2) geochemical synthesis in the ocean. The first scenario is well-established based on numerous measurements of amino acids in primitive solar system materials, such as in carbonaceous chondrites (Kvenvolden et al., 1970, Cronin & Pizzarello, 1997), and the coma of comet 67P/Churyumov-Gerasimenko (Altwegg et al., 2016). The classic mechanism for forming meteoritic amino acids is the Strecker synthesis (e.g., Elsila et al., 2016), as this process requires liquid water and numerous carbonaceous chondrites contain mineralogical evidence of aqueous alteration (Brearley, 2006). Cometary glycine could have been formed by ultraviolet irradiation of interstellar ices (Bernstein et al., 2002, Muñoz Caro et al., 2002). The plausible presence of abundant primordial amino acids in these moons building block materials suggests a simple first hypothesis for the origin of amino acids in Enceladus, Europa, and other small ocean worlds: that is, these bodies could have acquired primordial amino acids in an analogous manner to carbonaceous chondrites and comets. For the second scenario, serpentinization reactions have been demonstrated both in the lab and the field to promote abiotic synthesis of diverse complex organic compounds including hydrocarbons, carboxylic acids, amino acids, etc. by the reduction of inorganic precursors (Huber & Wächtershäuser, 1998, Huber et al., 2003, Proskurowski et al., 2008, Novikov & Copley, 2013). Recently, in-situ evidence of abiotically formed tryptophan was obtained at depth beneath the Atlantis Massif (Mid-Atlantic Ridge), suggesting that the serpentinizing hydrothermal field could synthesize some prebiotic molecules of interest efficiently (Ménez et al., 2018).

This paper assesses the first scenario of the plausibility of primordial amino acids surviving in the oceans of icy worlds, such as those in Enceladus and Europa. Could amino acids derived from primordial or other processes in the distant past persist until today? Amino acids are susceptible to destruction reactions, and the rate of destruction depends on the structure and side-chain properties of the amino acid. This affords an opportunity to identify highly reactive amino acids whose presence may imply geologically recent synthesis. Such a scenario would be consistent with the presence of active biota, or at least an abiotic (e.g., hydrothermal) geochemistry that is capable of making some of the molecules of life – both exciting possibilities. Here, we use published chemical kinetics data to estimate decomposition timescales of amino acids in liquid water for scenarios that may be relevant to Enceladus and Europa.

## 2. RATE CONSTANT CALCULATIONS

The decomposition of amino acids in water has been extensively studied for its applications to organic geochemistry (e.g. Ito et al., 2006, 2009), extracting and recycling proteinaceous products in industry or biotechnology (e.g., Zhu et al., 2011, Changi et al., 2015) and in investigations of the hydrothermal origin of life (e.g. Bada et al., 1995). Below the critical point of water (374°C, 22.1 MPa), the major reaction pathways of decomposition are decarboxylation to amines, and deamination to carboxylic acids (e.g., Cox & Seward, 2007, Li & Brill, 2003a, b). A third pathway, dehydration, can occur for OH-bearing amino acids such as serine and threonine (Li & Brill, 2003a).

In this study, we collected literature kinetics data for the decomposition of individual amino acids in liquid water over available ranges of temperature. The observed rate constant reflects the contributions of all reaction pathways as described above and follows the pseudo-first-order rate law (Li & Brill, 2003a) shown below

$$[AA] = [AA]_0 \exp(-kt) \tag{1}$$

which indicates that the concentration of amino acid decays exponentially with time *t*. Published values of rate constant *k* or ln *k* at a given temperature were used, or *k* was calculated from the amino acid decomposition half-life or *e*-folding lifetime (0.63-life, the time over which the concentration decreases to 1/*e* of its initial value). The rate constant depends on the conditions of the corresponding experiment. Data were obtained for fourteen amino acids: glycine, alanine, valine, leucine, isoleucine, methionine, serine, threonine, proline, aspartic acid, phenylalanine, histidine, arginine, and glutamic acid. The kinetic data and their references are available in Supplementary Table S1.

In order to predict rate constants at temperatures relevant to liquid water environments on ocean worlds (e.g., 0-374°C), all the rate constants that we found in the literature for each amino acid were used to perform least squares fitting and extrapolation to the functional form

$$\ln k = \ln A - \frac{E_a}{RT} \qquad (2)$$

where $A$ is the frequency factor, $E_a$ is the activation energy, $T$ is the absolute temperature, and $R$ is the gas constant (8.314 J mol$^{-1}$ K$^{-1}$). As in many previous studies (e.g., Vallentyne, 1964, Andersson & Holm, 2000, Abdelmoez et al., 2007), all of the rate constants used here were found to be consistent with the Arrhenius relationship of Eq. (2). Fig. 1 shows these relationships in which the mean value for the intercept (α) and slope coefficient (β) is (ln *A)* and *(-E$_a$/R)* respectively. For $\varepsilon_i$ is the residual error for each data point $x_i$, $y_i$ ($x_i = 1/T_i$ and $y_i = \ln k_i$) (Eq. 3) and n-2 indicates degrees of freedom, the standard errors $s_{\bar{\beta}}$ and $s_{\bar{\alpha}}$ were estimated by student-t distribution approach with confidence level of 95% (4, 5). Using this approach, we can construct a confidence interval for the slope β and the intercept α (Eq. 6, 7, Fig. 2). The derived values of $E_a$ and ln *A* and corresponding standard errors are summarized in Table 1.

$$\varepsilon_i = y_i - bx_i - a \qquad (3)$$

$$s_{\bar{\beta}} = \sqrt{\frac{\frac{1}{n-2}\sum_{i=1}^{n}\varepsilon_i^2}{\sum_{i=1}^{n}(x_i-\bar{x})^2}} \qquad (4)$$

$$s_{\bar{\alpha}} = \sqrt{\frac{1}{n(n-2)} \left(\sum_{i=1}^{n} \varepsilon_i^2\right) \frac{\sum_{i=1}^{n} x_i^2}{\sum_{i=1}^{n}(x_i - \bar{x})^2}} \tag{5}$$

$$\beta \in [\bar{\beta} - s_{\bar{\beta}} t^*_{n-2}, \bar{\beta} + s_{\bar{\beta}} t^*_{n-2}] \tag{6}$$

$$\alpha \in [\bar{\alpha} - s_{\bar{\alpha}} t^*_{n-2}, \bar{\alpha} + s_{\bar{\alpha}} t^*_{n-2}] \tag{7}$$

- Figure 1 here (submitted separately)

**Figure 1.** Least-squares fitting of the natural log of the rate constant vs. inverse temperature for the experimentally observed decomposition of each amino acid in water. Literature sources for these data are described in the main text.

- Figure 2 here (submitted separately)

**Figure 2.** Calculated rate constants at 273-600 K. The predicted rate constants vs. temperature lies within the 95% confidence interval (grey shaded area).

| Amino acid | Abbrv | $E_a$ (kJ mol$^{-1}$) | ln $A$ (s$^{-1}$) | Temperature range (K) |
|---|---|---|---|---|
| Glycine | Gly | 196 ± 12 | 36.7 ± 2.7 | 473-603 |
| Alanine | Ala | 149.0 ± 8.8 | 26.0 ± 1.9 | 473-603 |
| Valine | Val | 189 ± 16 | 33.9 ± 3.5 | 489-603 |
| Leucine | Leu | 146 ± 19 | 25.6 ± 4.3 | 473-603 |
| Isoleucine | Ile | 117.1± 12.2 | 19.5±2.6 | 503-603 |
| Methionine | Met | 168 ± 24 | 32.1±5.2 | 503-573 |
| Phenylalanine | Phe | 169 ± 12 | 28.9±3.0 | 450-603 |
| Serine | Ser | 147.0 ± 8.6 | 28.7±2.0 | 425-573 |
| Aspartic acid | Asp | 111.1±15.4 | 22.8±3.7 | 473-573 |
| Threonine | Thr | 130 ± 6 | 25.0±1.4 | 386-573 |
| Proline | Pro | 172 ± 31 | 28.5±6.78 | 489-613 |
| Histidine | His | 156±14 | 27.9±3.1 | 503-603 |
| Arginine* | Arg | 83.2 | 11.7 | 374-489 |
| Glutamic acid** | Glu | 152.2±6.8 | 21.7±1.6 | 489-553 |

**Table 1.** Amino acid decomposition rate constants derived from Fig. 1. *Parameters from *Vallentyne, 1968* **Parameters from *Povoledo & Vallentyne, 1964*.

## 3. MODELING THE DECOMPOSITION OF AMINO ACIDS

*3.1. Model framework and constraints*

According to Eq. (1), an amino acid would never totally disappear in theory, but we can obtain a practical constraint for that endmember scenario of destruction by defining the decomposition timescale for each amino acid as the 0.999-life (i.e., the time after which 99.9% of the initial concentration would be destroyed by a first-order process). As an example, this constraint would be consistent with an individual amino acid concentration of 1 µM decreasing below the detection limit (1 nM) recommended for mission success by the Europa Lander Science Definition Team (Hand et al., 2017). A similar approach with the lifetime concept was used to estimate racemization timescales for meteoritic amino acids based on asteroidal parent body temperatures (Cohen & Chyba, 2000). We solve analytically the integrated first-order rate equation (Eq. 1) with the corresponding kinetic rate constant calculated using Eq. (2) to quantify the destruction timescales.

Hydrothermal circulation is an important consideration for icy worlds that have an ocean that is in contact with a rocky interior (e.g., Enceladus, Europa), because amino acids decompose much faster at elevated temperatures (Fig. 2). This means that the decomposition timescale of amino acids depends strongly on 1) the maximum temperature experienced during hydrothermal circulation, 2) the total duration of amino acid exposure to hydrothermal temperatures and 3) the relative tendency of these amino acids to adsorb onto minerals. We need to know the total length of time that a fluid parcel has been in a hydrothermal system. This can involve multiple cycles of the fluid through the rocky interior over the history of the body. The total hydrothermal duration can be constrained based on 1) the mean residence time of water in the entire core, 2) the hydrothermal temperature of the ocean water when cycling through the core and 3) the timescale that hydrothermal activity could be sustained. Below, we focus on Enceladus, the icy world for which we have the most data.

For the first constraint, the derived moment of inertia factor of Enceladus (Iess et al., 2014, McKinnon, 2015) indicates a low-density core, which can be explained if there is considerable water-filled porosity (Choblet et al., 2017, Waite et al., 2017). We consider flows of water between two connected reservoirs: a flow of water from the ocean to the core, and a reverse flow from the core to the ocean. Assuming that the system is at steady state, the rates of exchange (*r*) between these reservoirs would be equal resulting in the condition

$$r = \frac{m_{ow}}{\tau_{ow}} = \frac{m_{pw}}{\tau_{pw}} \qquad (8)$$

where *m* is the mass of ocean water or pore water, and *τ* is the residence time per cycle of a water parcel between the reservoirs. The relationship between the individual residence times and the total time that hydrothermal activity is sustained in the core (*L*) can be expressed as

$$L = \omega(\tau_{ow} + \tau_{pw}) \qquad (9)$$

where *ω* is the number of cycles over the lifetime of the ocean. By combining Eqs. (8) and (9), we can obtain the following expression for the total duration (*d*) that a fluid spends in the core

$$d = \omega\tau_{pw} = L(1 + \frac{m_{ow}}{\tau_{ow}})^{-1} \qquad (10)$$

It is evident that the fraction of time that a fluid spends in the core depends on the relative masses of water in the two reservoirs. The masses of ocean water and pore water on Enceladus have been estimated to be ~2-3×10$^{19}$ kg (Thomas et al., 2016, Čadek et al., 2016) and 8×10$^{18}$ kg (Waite et al. 2017), respectively. If $m_{ow}/m_{pw}$ is approximated as constant through time, then the total time that a water parcel may have resided in the core would occupy a range of 20 to 40 percent of the age of Enceladus' ocean.

The temperature of ocean water cycling through the core has been constrained based on the temperature required for the formation of certain chemical species erupted from Enceladus: 1) nanosilica particles, which are thought to form at the ocean floor of Enceladus where $SiO_2$-enriched hydrothermal fluid mixes with the cold seawater (Hsu et al., 2015); and 2) anomalously high abundance of native molecular $H_2$ that could not be produced in such quantities by other sources (Waite et al. 2017, Glein et al., 2018). A summary table of the minimum required temperature for the formation of hydrothermal species for different scenarios is given below (Table 2).

Clearly, there are a number of possible hydrothermal temperatures, but we consider a temperature above ~170°C to be most plausible, as it is consistent with constraints from both $H_2$ and silica. It was suggested that amorphous silica could precipitate from hydrothermal fluids as cool as ~50°C (Sekine et al., 2015), but such temperatures may not be conducive to making sufficient $H_2$ to explain the observation (Glein et al., 2018). If silica and $H_2$ are formed in different types/regions of hydrothermal systems, then any amino acids in the ocean would still undergo circulation through the hotter systems that produce $H_2$. Nevertheless, in an effort to err on the side of caution, we will also discuss our results for the case of a minimum hydrothermal temperature of 50°C.

| **Scenarios for the formation of hydrothermal species** | **Minimum required temperature (in Celsius & Kelvin)** | **References** |
|---|---|---|
| Constant hydrothermal fluid pH upon cooling to form nanosilica | 90°C (363K) | Hsu et al., 2015 |
| Hydrothermal fluid pH changes: Chemically closed, ocean-core system forming nanosilica | 150-200°C (423-473K) | Hsu et al., 2015, Sekine et al., 2015 |
| High pH hydrothermal fluid mixes with lower pH ocean water to form nanosilica | 50°C (323K) | Sekine et al., 2015 |
| The source of $SiO_2$ is quartz formed by carbonation of the seafloor environment | 165°C (438K) | Glein et al., 2018 |
| Formation of molecular $H_2$ by alteration of reduced rocks | 170°C (443K) | Glein et al., 2018 |

**Table 2**. The minimum required temperature for the formation of hydrothermal species for different scenarios.

The third constraint is the total time that hydrothermal activity is sustained in the core. The current preferred mechanism to sustain tidal dissipation is through the resonance-locking mechanism (Nimmo et al., 2018), which suggests that the equilibrium tidal heating varies only slowly with satellite semimajor axis, and hydrothermal activity could even persist for billions of years (Choblet et al., 2017). Here, we adopt a very conservative minimal hydrothermal activity timescale of 20 Myr calculated by Choblet et al. (2017) for the case of lateral variations of the ice shell in which direct contact might occur between the porous core and the ice in its thickest equatorial regions, at the sub-Saturnian and anti-Saturnian points. Combining with the first constraint on the total duration (*d*) that a fluid spends in the core, the total duration of amino acid subjugation to hydrothermal temperatures would be in the range of 4-8 Myr. In the most long-lasting scenario in which the ocean could sustain hydrothermal activity for billions of years since its formation (Choblet et al., 2017), the total duration of amino acid exposure to hydrothermal temperatures could be up to 1.8 Gyr.

The relative tendency of these amino acids to adsorb/concentrate on minerals is another factor that would affect their reactivity and what suite of species could be detected. It is possible that not all primordially formed amino acids can be extracted into the aqueous phase for decomposition in water instead of being adsorbed/concentrated on mineral surfaces at the seafloor or in the rocky interior (Hazen & Sverjensky, 2010). Another possible implication for in-situ measurements is that newly formed species such as aspartic acid could be efficiently concentrated on layer double hydroxide minerals in an alkaline environment (Grégoire et al., 2016); however the precise effect is unknown for ocean worlds chiefly because we lack details concerning the mineralogy. Nonetheless, this complexity does not preclude the possibility that some of these species would be detectable in-situ.

*3.2. Model results*

Here, we present the decomposition timescale of all fourteen amino acids as a function of temperature (Fig. 3) and interpret the results for the endmember case that all primordial amino acids species could have accumulated in the ocean. In an icy moon in which the ocean is nearly at the freezing point of 273 K, the cold temperature would inhibit appreciable decomposition for most of the amino acids, except for aspartic acid and threonine (Table 3). The nominal value for the decomposition timescale of arginine at 273K also suggests that arginine may not persist over a geological timescale. However, more experimental rate constants are critically needed to estimate the lower and upper bound of the decomposition timescale and therefore, the fate of arginine in a cold ocean scenario. The other amino acids could persist in water at 273 K over the age of the Solar System, and their survival would be enhanced in colder oceans maintained by antifreezes such as ammonia or salts. This finding suggests that some amino acids on non-hydrothermally active bodies (if such worlds exist) can be primordial.

Because amino acids are more reactive at a higher temperature, a geologically short amount of time over which amino acids are subjected to a moderate hydrothermal temperature of 323K could lead to significant destruction of many amino acids. Many of them including alanine, arginine, histidine, isoleucine, leucine, methionine, serine should quickly decompose below 1nM in an ocean with a limited lifetime of liquid water (Fig. 3). On the other hand, the survivability (and therefore, the presence of primordial relics in the ocean) of some less reactive species such as glutamic acid, glycine, phenylalanine, proline, or valine would depend extensively on the duration that hydrothermal activity could be sustained in the ocean. If hydrothermal activity has persisted for billions of years since the moon's formation (Choblet et al., 2017), then all of the above species would likely be decomposed to concentrations below the nominal detection limit over multiple cycles through the hot rocky core.

In the nominal scenario for the temperature of hydrothermal systems, the interpretation completely changes when amino acids have experienced elevated temperatures of more than 443K even for a relatively short amount of time. All of these amino acids are predicted to decompose very quickly in a geological sense at temperatures greater than 443K, and completely disappear below 1 nM (with respect to the 0.999-life) in less than 1000 years, even for the least reactive species. As an illustrative example, the shaded gray area in Fig. 3 shows the confidence interval of our simulated decomposition timescale for glycine – one of the most resistant species to decomposition. The result indicates that even for the longest decomposition timescale constrained by the available literature rate constants, amino acids could decompose in a time short compared to geologic time, upon multiple cycles through a hydrothermal rocky core.

All fourteen amino acids considered in this study have relatively short lifetimes at relevant hydrothermal temperatures. If we interpret these short lifetimes in the context of our knowledge of likely present-day hydrothermal activity at Enceladus (Hsu et al., 2015, Waite et al., 2017) and the strong potential for at least past hydrothermal activity at Europa (Hussmann & Spohn, 2004), then it is most plausible that any primordial amino acids at Enceladus or Europa have been destroyed. Thus, any amino acids now present must have formed in less than ~ 1 Myr ago. This interpretation is robust, especially for the most easily decomposed aspartic acid and threonine.

| Amino acid | Lower $t_{0.999}$ at 273K (yr) | Nominal $t_{0.999}$ at 273K (yr) | Upper $t_{0.999}$ at 273K (yr) |
|---|---|---|---|
| Alanine | $3.4 \times 10^8$ | $3.6 \times 10^{10}$ | $1.7 \times 10^{12}$ |
| Arginine | No reported error | $1.3 \times 10^4$ | No reported error |
| Aspartic acid | $1.0 \times 10^1$ | $4.9 \times 10^4$ | $1.0 \times 10^8$ |
| Glutamic acid | $8.5 \times 10^4$ | $1.1 \times 10^{13}$ | $5.4 \times 10^{20}$ |
| Glycine | $4.2 \times 10^{12}$ | $7.8 \times 10^{14}$ | $6.4 \times 10^{16}$ |
| Histidine | $1.7 \times 10^7$ | $1.1 \times 10^{11}$ | $3.3 \times 10^{14}$ |
| Isoleucine | $5.8 \times 10^3$ | $1.9 \times 10^7$ | $2.8 \times 10^{10}$ |
| Leucine | $2.9 \times 10^4$ | $1.2 \times 10^{10}$ | $1.7 \times 10^{15}$ |
| Methionine | $3.9 \times 10^5$ | $3.7 \times 10^{11}$ | $1.6 \times 10^{17}$ |
| Phenylalanine | $2.2 \times 10^{10}$ | $1.2 \times 10^{13}$ | $3.1 \times 10^{15}$ |
| Proline | $1.8 \times 10^6$ | $6.7 \times 10^{13}$ | $1.1 \times 10^{21}$ |
| Serine | $1.4 \times 10^7$ | $9.9 \times 10^8$ | $5.1 \times 10^{10}$ |
| Threonine | $1.2 \times 10^6$ | $2.7 \times 10^7$ | $2.8 \times 10^8$ |
| Valine | $4.6 \times 10^{10}$ | $5.4 \times 10^{14}$ | $2.9 \times 10^{18}$ |

**Table 3.** Nominal, lower and upper bound values for the decomposition timescale of the considered amino acids at 273K. The lower and upper bounds were calculated based on the standard errors of rate constants given in **Table 1**. Similarly, **Fig. 2** shows the 95% confidence interval of the extrapolated rate constants in natural log scale (shaded grey area).

- Figure 3 here (submitted separately)

**Figure 3.** Decomposition timescale of amino acids in aqueous solution. The cyan lines indicate bounds on the exposure time of amino acids to elevated temperatures in the rocky core. The horizontal dash line shows the duration that amino acids are subjected to hydrothermal temperature in the most long-lasting scenario in which the ocean could sustain hydrothermal activity for billions of years since its formation. The vertical dotted lines correspond to nominal and conservative lower limits on those temperatures (Glein et al., 2018; Sekine et al., 2015).

## 4. DISCUSSION

The underlying assumption of our analysis is that the rate constants follow first-order kinetics. This is appropriate unless the daughter products exert an autocatalytic effect on the mechanism of the reaction (Li and Brill 2003a). In this case, the first-order rate law would apply during an induction period (at least to 40% conversion), but later as the process continues, the concentration vs. time profile would follow a second-order rate constant. The consequence would be faster amino acid destruction, which would bolster the argument of present amino acids not being primordial and formed even more recently. Another simplification is that the calculations at low temperatures use extrapolated rate constants, as experimental data only exist at high temperatures (374-613 K) where decomposition can be monitored over the timescale of laboratory experiments. The development of mechanistic models for each decomposition pathway that incorporate transition state theory and account for changes in solvation with temperature would be helpful, as such models would have a firm underpinning for extrapolation to low temperatures. The strong dependence of amino acids decomposition timescales on the maximum temperature to which they are subjected suggests that tighter constraints on temperatures of core fluids are needed and could be obtained using tools of geothermometry, e.g. isotopic fractionation between simple volatiles, ratios of certain organic compounds, or "clumping" of rare isotopes (Glein et al., 2018).

The next generation of geochemical models also needs to account for the effects of additional physiochemical parameters beyond temperature. As amino acids decompose in aqueous solution, their rate constants would vary with pH as well. On ocean worlds, the pH will be controlled by the balance of inputs of alkalinity from water-rock reactions (Glein et al., 2015), and acidity from oxyanion formation (Pasek & Greenberg, 2012, Ray et al., 2017). If the pH is not far from neutral, then the pH effect may be minor. In fact, the observed rate constants are relatively independent of the solution pH in the range 3-8.5 (the range in which the dominant amino acid form is the dipolar zwitterion) for many amino acids, namely glycine, alanine, valine, leucine, isoleucine, serine, threonine, methionine, and proline (Li & Brill 2003, a, b). Other factors may include the effects of certain minerals, including catalysis. Norvaline, a representative of alkyl-α-amino acids, decomposes more than 10 times faster in the presence of the mineral assemblage pyrite-pyrrhotite-magnetite (PPM) than in its absence (McCollom, 2013). Minerals could also alter the oxidation state and indirectly affect the stability of amino acids, e.g. abundant $H_2$ in hydrothermal fluids is known to slow down the rates of some decomposition pathways that produce $H_2$ (Andersson & Holm, 2000, Lee et al., 2014). However, it is unclear if amino acids can persist metastably under such conditions in dynamic systems where $H_2$ is lost to space, the loss of which progressively oxidizes the water-rock system. Fully coupled kinetic-thermodynamic-transport models of heterogeneous phase chemistry are needed to better understand the behavior of amino acids and other organics in the complex environment. For in-situ measurements, it is also important to note that the detection in the aqueous phase will be affected by the relative tendency of these amino acids to adsorb onto minerals at the seafloor, and it may be difficult to detect such a process directly.

## 5. CONCLUSION

By examining the decomposition timescale of amino acids in aqueous solution based on existing laboratory kinetics data, we found that all fourteen amino acids considered in this study decompose to a great extent in relatively short amounts of time in hydrothermally active oceans. The short lifetimes of amino acids suggest that if they are detected with a concentration above 1 nM on Enceladus, Europa, and other hydrothermally active ocean worlds, their origin should come from active production rather than primordial synthesis. One could argue that cometary impacts onto the ice might replenish the primordial population, however, projectile impact escape is dominant on icy moons, given the escape velocity of 2 km/s on Europa (and nearly an order of magnitude lower for Enceladus), so cometary impacts would provide a negligible contribution to the organics reservoir (Pierazzo & Chyba, 2006). In particular, the detection of aspartic acid, and threonine would provide the strongest evidence for an active production of amino acids within the ocean. None of these species can be primordial and so in-situ detection would indicate recent production of the amino acid (<1 Myr), via geochemical or biotic pathways. And finally, as the decomposition timescale of amino acids is relatively short, even shorter than the timescale of oceans existing on Enceladus and Europa, models considering the production of amino acids should take into account the role of the decomposition process as a necessary component of the (bio)geochemistry of amino acids on ocean worlds.

**ACKNOWLEDGEMENTS**

The authors acknowledge the contribution of E. Shock and helpful discussions regarding residence times in reservoirs of different sizes with E. Kite and S. Desch in the original calculations reported in Monroe et al., 2017, which inspired the present work. Support from the Cassini Project and a JPL Distinguished Visiting Scientist position (JIL) are gratefully acknowledged.

**AUTHOR DISCLOSURE STATEMENT**

No competing financial interests exist.

Lemke, K. H., Rosenbauer, R. J., & Bird, D. K. (2009). Peptide Synthesis in Early Earth Hydrothermal Systems. *Astrobiology*, *9*(2), 141–146. https://doi.org/10.1089/ast.2008.0166

Li, J., & Brill, T. B. (2003a). Spectroscopy of Hydrothermal Reactions 25: Kinetics of the Decarboxylation of Protein Amino Acids and the Effect of Side Chains on Hydrothermal Stability. *The Journal of Physical Chemistry A*, *107*(31), 5987–5992. https://doi.org/10.1021/jp0224766

Li, J., & Brill, T. B. (2003b). Spectroscopy of hydrothermal reactions, part 26: Kinetics of decarboxylation of aliphatic amino acids and comparison with the rates of racemization. *International Journal of Chemical Kinetics*, *35*(11), 602–610. https://doi.org/10.1002/kin.10160

Lowell, R. P., & DuBose, M. (2005). Hydrothermal systems on Europa. *Geophysical Research Letters*, *32*(5). https://doi.org/10.1029/2005GL022375

Lunine, J. I. (2017). Ocean worlds exploration. *Acta Astronautica*, *131*, 123–130. https://doi.org/10.1016/j.actaastro.2016.11.017

Martin, W., Baross, J., Kelley, D., & Russell, M. J. (2008). Hydrothermal vents and the origin of life. *Nature Reviews Microbiology*, *6*(11), 805–814. https://doi.org/10.1038/nrmicro1991

McCollom, T. M. (2013). The influence of minerals on decomposition of the n-alkyl-α-amino acid norvaline under hydrothermal conditions. *Geochimica et Cosmochimica Acta*, *104*, 330–357. https://doi.org/10.1016/j.gca.2012.11.008

McKinnon, W. B. (2015). Effect of Enceladus's rapid synchronous spin on interpretation of Cassini gravity: Enceladus: global ocean, thick ice shell. *Geophysical Research Letters*, *42*(7), 2137–2143. https://doi.org/10.1002/2015GL063384
20

Porco, C., DiNino, D., & Nimmo, F. (2014). How the Geysers, Tidal Stresses, and Thermal Emission across the South Polar Terrain of Enceladus are Related. *The Astronomical Journal*, *148*(3), 45. https://doi.org/10.1088/0004-6256/148/3/45

Postberg, F., Schmidt, J., Hillier, J., Kempf, S., & Srama, R. (2011). A salt-water reservoir as the source of a compositionally stratified plume on Enceladus. *Nature*, *474*(7353), 620–622. https://doi.org/10.1038/nature10175

Postberg, Frank, Khawaja, N., Abel, B., Choblet, G., Glein, C. R., Gudipati, M. S., … Waite, J. H. (2018). Macromolecular organic compounds from the depths of Enceladus. *Nature*, *558*(7711), 564–568. https://doi.org/10.1038/s41586-018-0246-4

Povoledo, D., & Vallentyne, J. R. (1964). Thermal reaction kinetics of the glutamic acid-pyroglutamic acid system in water. *Geochimica et Cosmochimica Acta*, *28*(5), 731–734. https://doi.org/10.1016/0016-7037(64)90089-4

Proskurowski, G., Lilley, M. D., Seewald, J. S., Früh-Green, G. L., Olson, E. J., Lupton, J. E., … Kelley, D. S. (2008). Abiogenic Hydrocarbon Production at Lost City Hydrothermal Field. *Science*, *319*(5863), 604–607. https://doi.org/10.1126/science.1151194

Ray, C., Waite, J. H., Jr., Glein, C., & Teolis, B. D. (2017). Oxidation in Enceladus' Ocean. *AGU Fall Meeting Abstracts*, *51*. Retrieved from http://adsabs.harvard.edu/abs/2017AGUFM.P51F..08R

Roth, L., Saur, J., Retherford, K. D., Strobel, D. F., Feldman, P. D., McGrath, M. A., & Nimmo, F. (2014). Transient Water Vapor at Europa's South Pole. *Science*, *343*(6167), 171–174. https://doi.org/10.1126/science.1247051

Russell, M. J., & Hall, A. J. (2006). The onset and early evolution of life. In *Memoir 198: Evolution of Early Earth's Atmosphere, Hydrosphere, and Biosphere - Constraints from*
22

# Table S1. Literature Rate Constants

| Amino acid | Temperature (K) | Rate constant (s$^{-1}$) | References |
|---|---|---|---|
| Alanine | 603 | 3.70E-02 | Li et al., 2002 |
| | 603 | 1.80E-02 | Li & Brill, 2003b |
| | 593 | 1.02E-02 | Li & Brill, 2003b |
| | 593 | 1.80E-02 | Li et al., 2002 |
| | 583 | 5.00E-03 | Li & Brill, 2003b |
| | 583 | 9.10E-03 | Li et al., 2002 |
| | 573 | 4.72E-03 | Li et al., 2002 |
| | 573 | 1.93E-02 | Sato et al., 2004 |
| | 563 | 2.20E-03 | Li et al., 2002 |
| | 563 | 1.02E-03 | Abdelmoez et al., 2007 |
| | 553 | 1.11E-03 | Li et al., 2002 |
| | 533 | 3.33E-04 | Abdelmoez et al., 2007 |
| | 503 | 9.50E-05 | Abdelmoez et al., 2007 |
| | 473 | 7.11E-06 | Andersson & Holm, 2000 |
| | 473 | 4.70E-06 | Snider & Wolfenden, 2000 |
| Arginine | 489 | 5.38E-04 | Vallentyne 1968 |
| | 374 | 3.31E-07 | Vallentyne 1968 |
| Aspartic acid | 573 | 4.80E-01 | Sato et al., 2004 |
| | 533 | 2.30E-01 | Faisal et al., 2007 |
| | 513 | 3.40E-02 | Faisal et al., 2007 |
| | 503 | 1.20E-02 | Abdelmoez et al., 2007 |
| | 493 | 1.10E-02 | Faisal et al., 2007 |
| | 473 | 3.00E-03 | Faisal et al., 2007 |
| | 473 | 8.60E-03 | Andersson & Holm, 2000 |
| Glutamic acid | 553 | 1.08E-05 | Povoledo & Vallentyne, 1964 |
| | 525 | 2.21E-06 | Povoledo & Vallentyne, 1964 |
| | 489 | 1.45E-07 | Povoledo & Vallentyne, 1964 |
| Glycine | 603 | 9.50E-02 | Li & Brill, 2003b |
| | 593 | 6.20E-02 | Li & Brill, 2003b |
| | 583 | 3.70E-02 | Li & Brill, 2003b |
| | 573 | 1.93E-02 | Sato et al., 2004 |
| | 563 | 2.33E-03 | Abdelmoez et al., 2007 |
| | 533 | 2.33E-04 | Abdelmoez et al., 2007 |
| | 503 | 3.00E-05 | Abdelmoez et al., 2007 |
| | 473 | 3.60E-06 | Snider & Wolfenden, 2000 |
| Histidine | 603 | 4.53E-02 | Li & Brill, 2003a |

|  | 593 | 3.03E-02 | Li & Brill, 2003a |
|  | 583 | 1.69E-02 | Li & Brill, 2003a |
|  | 573 | 9.28E-03 | Li & Brill, 2003a |
|  | 563 | 2.33E-03 | Abdelmoez et al., 2007 |
|  | 533 | 3.67E-04 | Abdelmoez et al., 2007 |
|  | 503 | 1.43E-04 | Abdelmoez et al., 2007 |
| Isoleucine | 603 | 2.89E-02 | Li & Brill, 2003b |
|  | 593 | 1.68E-02 | Li & Brill, 2003b |
|  | 583 | 8.08E-03 | Li & Brill, 2003b |
|  | 563 | 2.50E+03 | Abdelmoez et al., 2007 |
|  | 533 | 6.50E-04 | Abdelmoez et al., 2007 |
|  | 503 | 3.00E-04 | Abdelmoez et al., 2007 |
| Leucine | 603 | 3.08E-02 | Li & Brill, 2003b |
|  | 593 | 1.95E-02 | Li & Brill, 2003b |
|  | 583 | 1.15E-02 | Li & Brill, 2003b |
|  | 573 | 3.97E-02 | Sato et al., 2004 |
|  | 563 | 2.10E-03 | Abdelmoez et al., 2007 |
|  | 555 | 2.92E-04 | Vallentyne 1964 |
|  | 533 | 2.00E-04 | Abdelmoez et al., 2007 |
|  | 503 | 6.33E-05 | Abdelmoez et al., 2007 |
|  | 473 | 2.43E-05 | Andersson & Holm, 2000 |
| Methionine | 573 | 4.47E-02 | Li & Brill, 2003a |
|  | 563 | 2.84E-02 | Li & Brill, 2003a |
|  | 563 | 1.05E-02 | Abdelmoez et al., 2007 |
|  | 553 | 1.79E-02 | Li & Brill, 2003a |
|  | 543 | 1.05E-02 | Li & Brill, 2003a |
|  | 533 | 1.47E-03 | Abdelmoez et al., 2007 |
|  | 503 | 3.33E-04 | Abdelmoez et al., 2007 |
| Phenylalanine | 603 | 2.86E-02 | Li & Brill, 2003a |
|  | 593 | 1.76E-02 | Li & Brill, 2003a |
|  | 583 | 9.07E-03 | Li & Brill, 2003a |
|  | 573 | 4.24E-03 | Li & Brill, 2003a |
|  | 568 | 2.37E-04 | Vallentyne 1964 |
|  | 563 | 1.17E-03 | Abdelmoez et al., 2007 |
|  | 562 | 2.16E-04 | Vallentyne 1964 |
|  | 556 | 1.58E-04 | Vallentyne 1964 |
|  | 552 | 1.14E-04 | Vallentyne 1964 |
|  | 546 | 7.10E-05 | Vallentyne 1964 |
|  | 539 | 4.97E-05 | Vallentyne 1964 |
|  | 533 | 2.67E-04 | Abdelmoez et al., 2007 |

|  | 526 | 3.10E-05 | Vallentyne 1964 |
|---|---|---|---|
|  | 505 | 9.04E-06 | Vallentyne 1964 |
|  | 503 | 4.83E-05 | Abdelmoez et al., 2007 |
|  | 498 | 5.52E-06 | Vallentyne 1964 |
|  | 483 | 1.77E-06 | Vallentyne 1964 |
|  | 482 | 2.48E-06 | Vallentyne 1964 |
|  | 481 | 1.17E-06 | Vallentyne 1964 |
|  | 450 | 2.07E-07 | Vallentyne 1964 |
| Proline | 613 | 7.17E-03 | Li & Brill, 2003a |
|  | 603 | 3.94E-03 | Li & Brill, 2003a |
|  | 593 | 2.05E-03 | Li & Brill, 2003a |
|  | 563 | 1.67E-04 | Abdelmoez et al., 2007 |
|  | 555 | 3.05E-05 | Vallentyne 1968 |
|  | 533 | 6.17E-05 | Abdelmoez et al., 2007 |
|  | 503 | 4.67E-05 | Abdelmoez et al., 2007 |
|  | 489 | 1.71E-07 | Vallentyne 1968 |
| Serine | 573 | 4.02E-01 | Sato et al., 2004 |
|  | 573 | 1.27E-01 | Li & Brill, 2003a |
|  | 563 | 7.94E-02 | Li & Brill, 2003a |
|  | 563 | 2.40E-02 | Abdelmoez et al., 2007 |
|  | 553 | 5.52E-02 | Li & Brill, 2003a |
|  | 543 | 3.41E-02 | Li & Brill, 2003a |
|  | 533 | 6.17E-03 | Abdelmoez et al., 2007 |
|  | 503 | 1.37E-03 | Abdelmoez et al., 2007 |
|  | 489 | 4.35E-04 | Vallentyne 1964 |
|  | 473 | 2.92E-04 | Andersson & Holm, 2000 |
|  | 458 | 2.13E-05 | Vallentyne 1964 |
|  | 425 | 4.55E-06 | Vallentyne 1964 |
| Threonine | 573 | 9.21E-02 | Li & Brill, 2003a |
|  | 563 | 5.39E-02 | Li & Brill, 2003a |
|  | 553 | 3.27E-02 | Li & Brill, 2003a |
|  | 543 | 1.76E-02 | Li & Brill, 2003a |
|  | 503 | 4.30E-03 | Abdelmoez et al., 2007 |
|  | 489 | 3.70E-04 | Vallentyne 1964 |
|  | 425 | 1.23E-05 | Vallentyne 1964 |
|  | 386 | 1.11E-07 | Vallentyne 1964 |
| Valine | 603 | 3.05E-02 | Li & Brill, 2003b |
|  | 593 | 1.60E-02 | Li & Brill, 2003b |
|  | 583 | 8.65E-03 | Li & Brill, 2003b |
|  | 563 | 1.02E-03 | Abdelmoez et al., 2007 |

|   |   |   |
|---|---|---|
| 533 | 4.50E-05 | Abdelmoez et al., 2007 |
| 503 | 1.83E-05 | Abdelmoez et al., 2007 |
| 489 | 5.78E-06 | Vallentyne 1964 |

**ADDITIONAL REFERENCES TO TABLE S1.**

# Figure 1

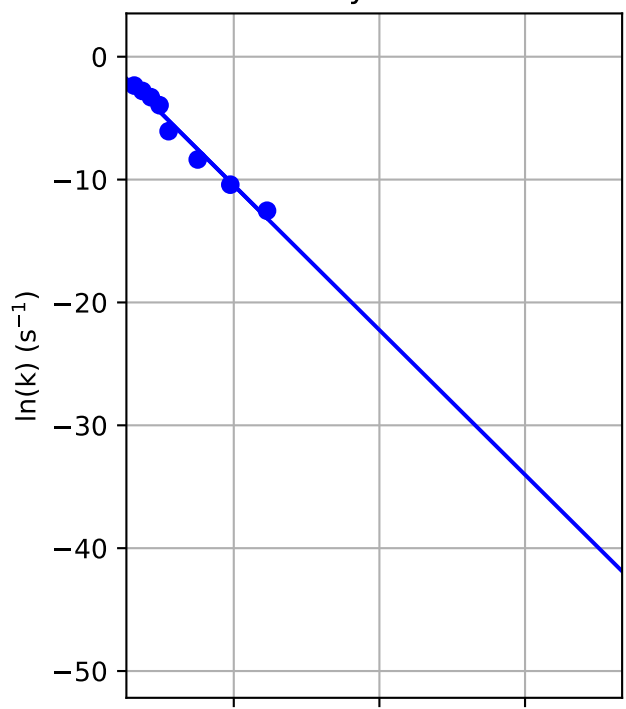
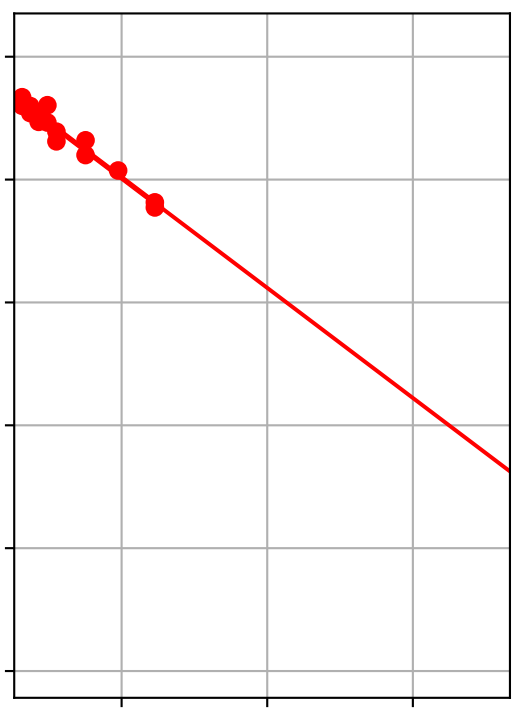
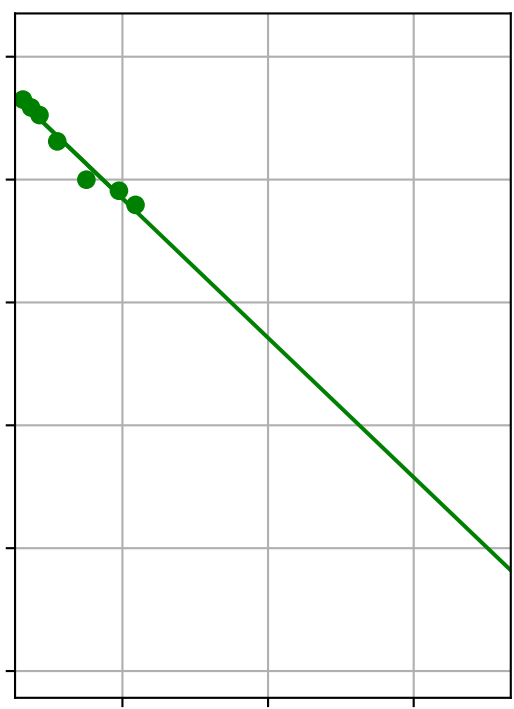
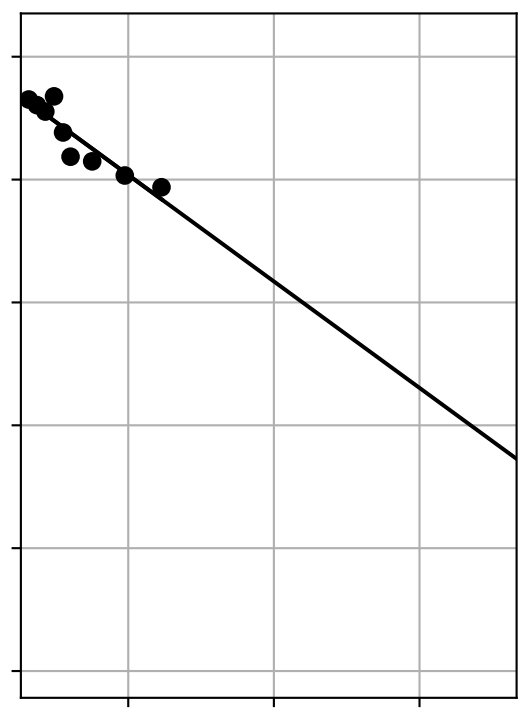
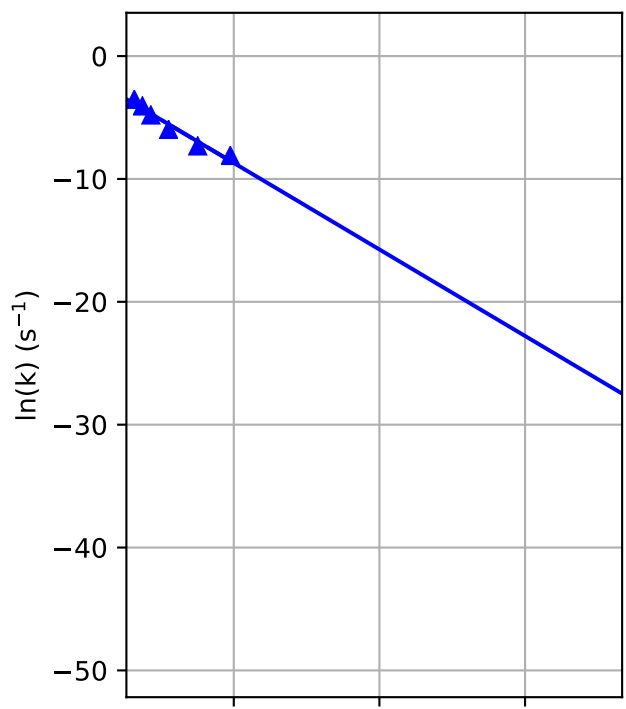
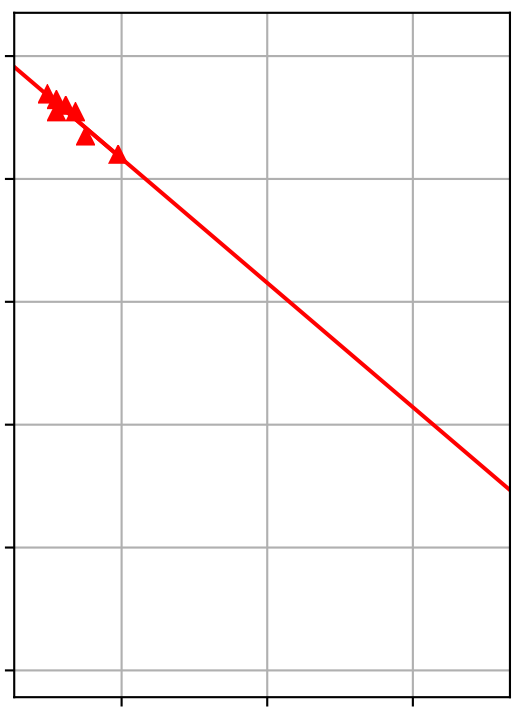
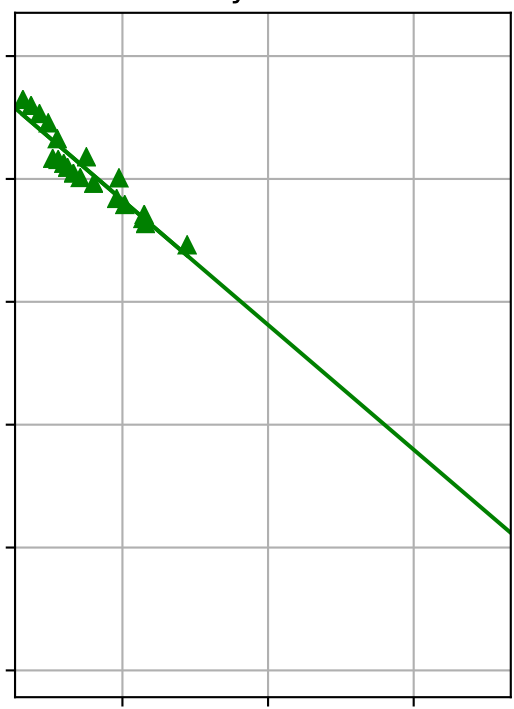
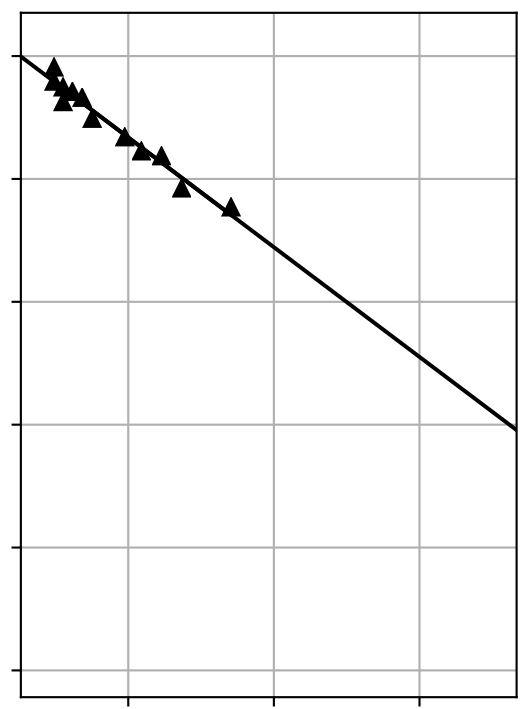
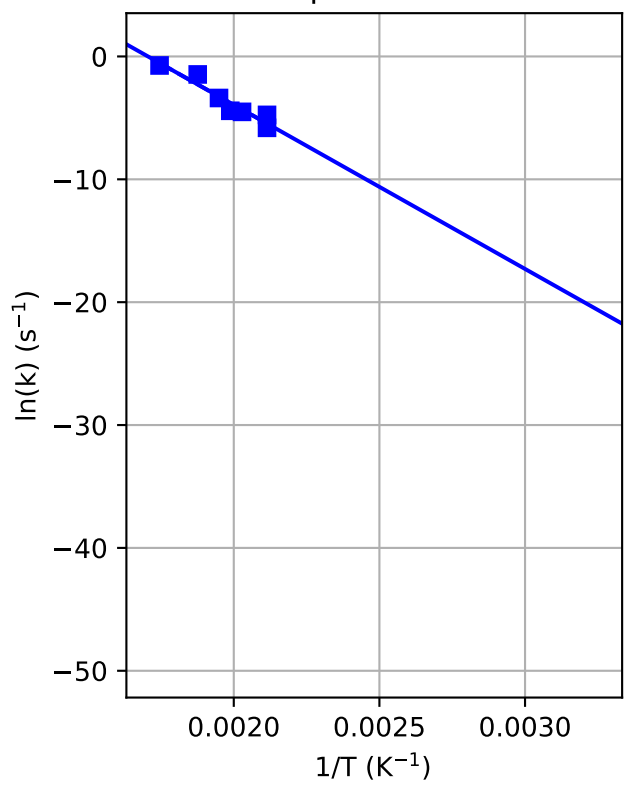
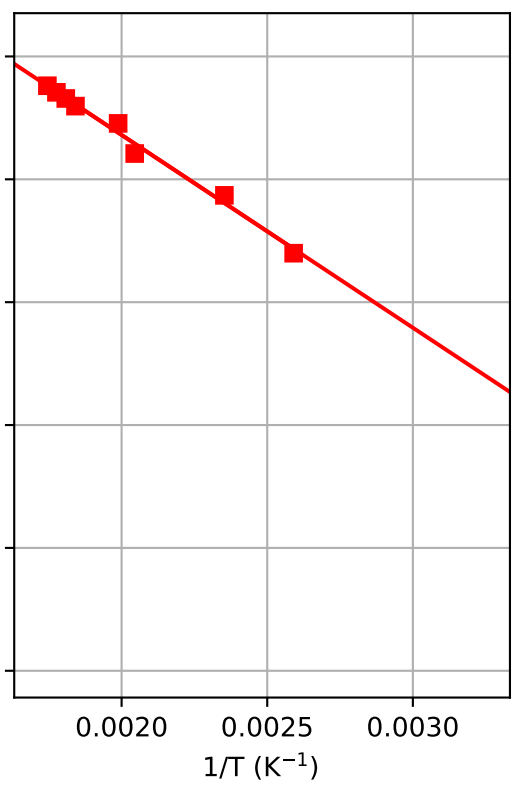
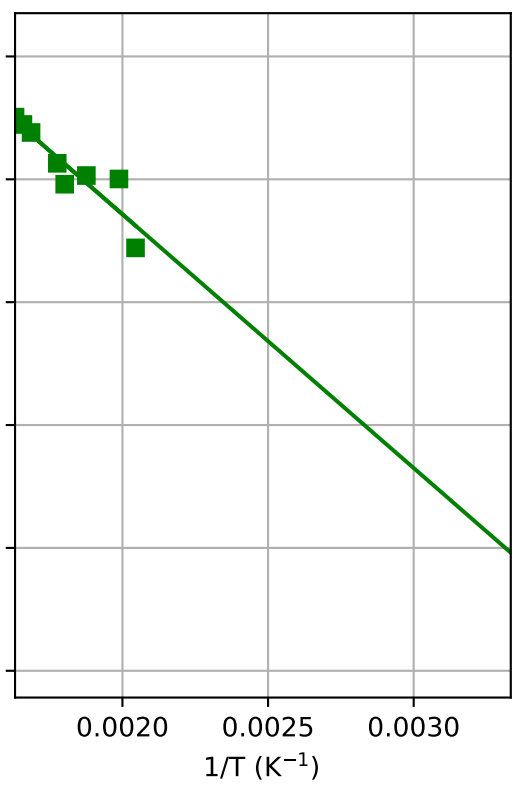
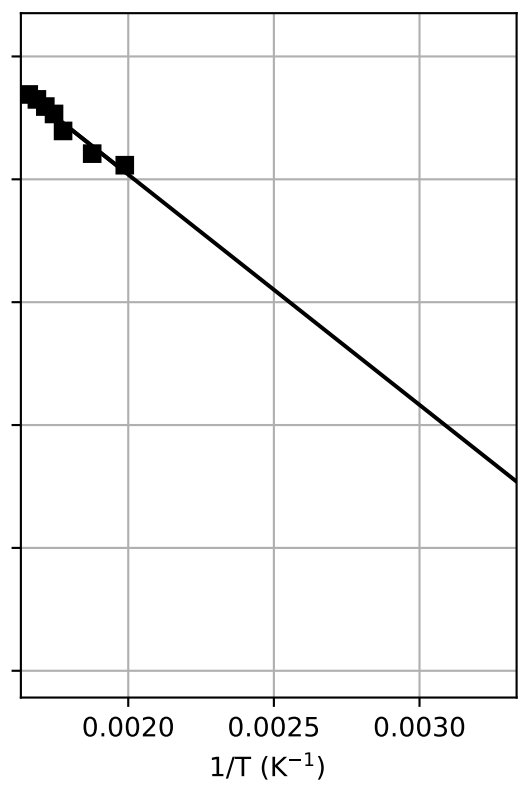

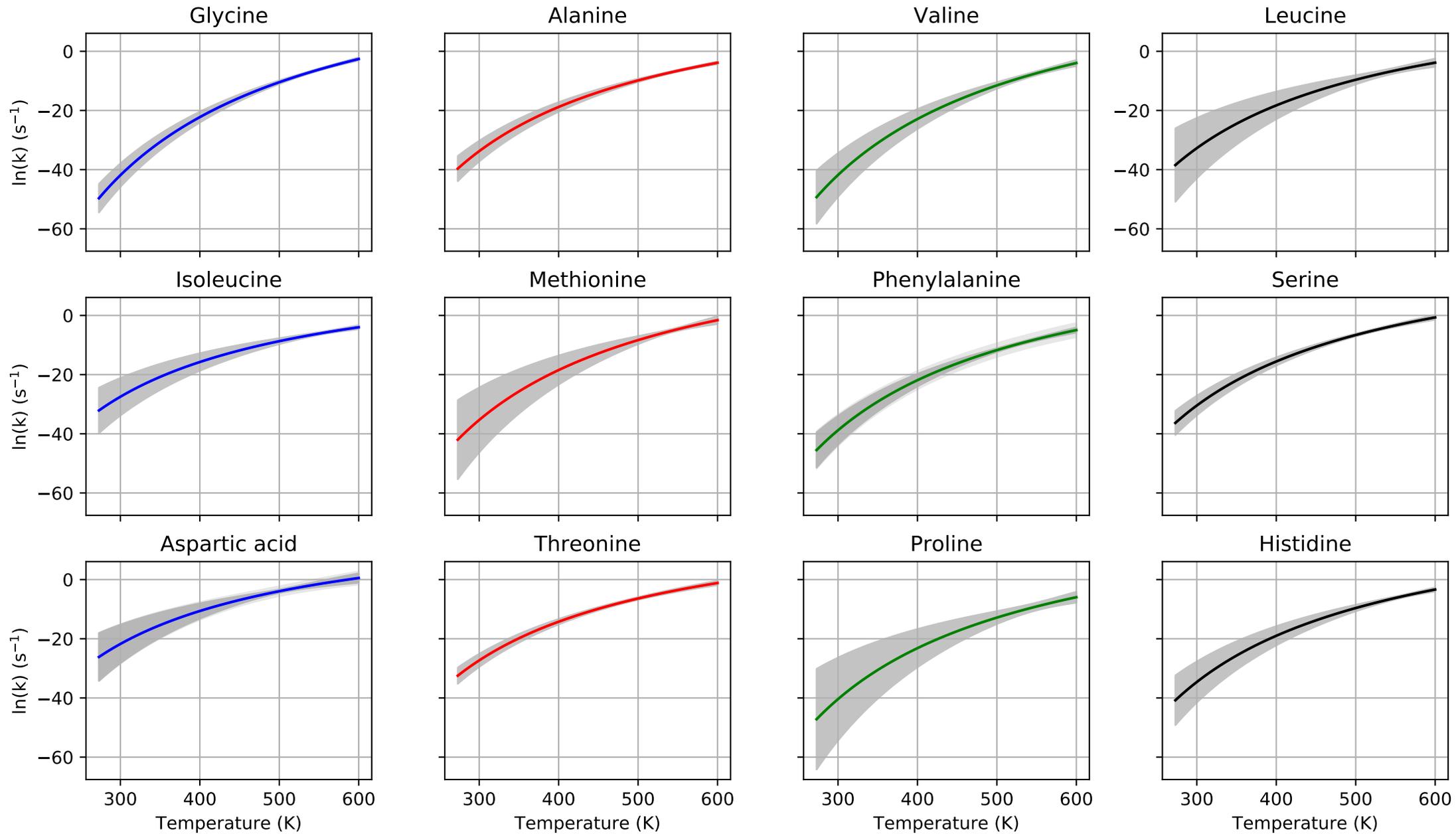

Figure 2

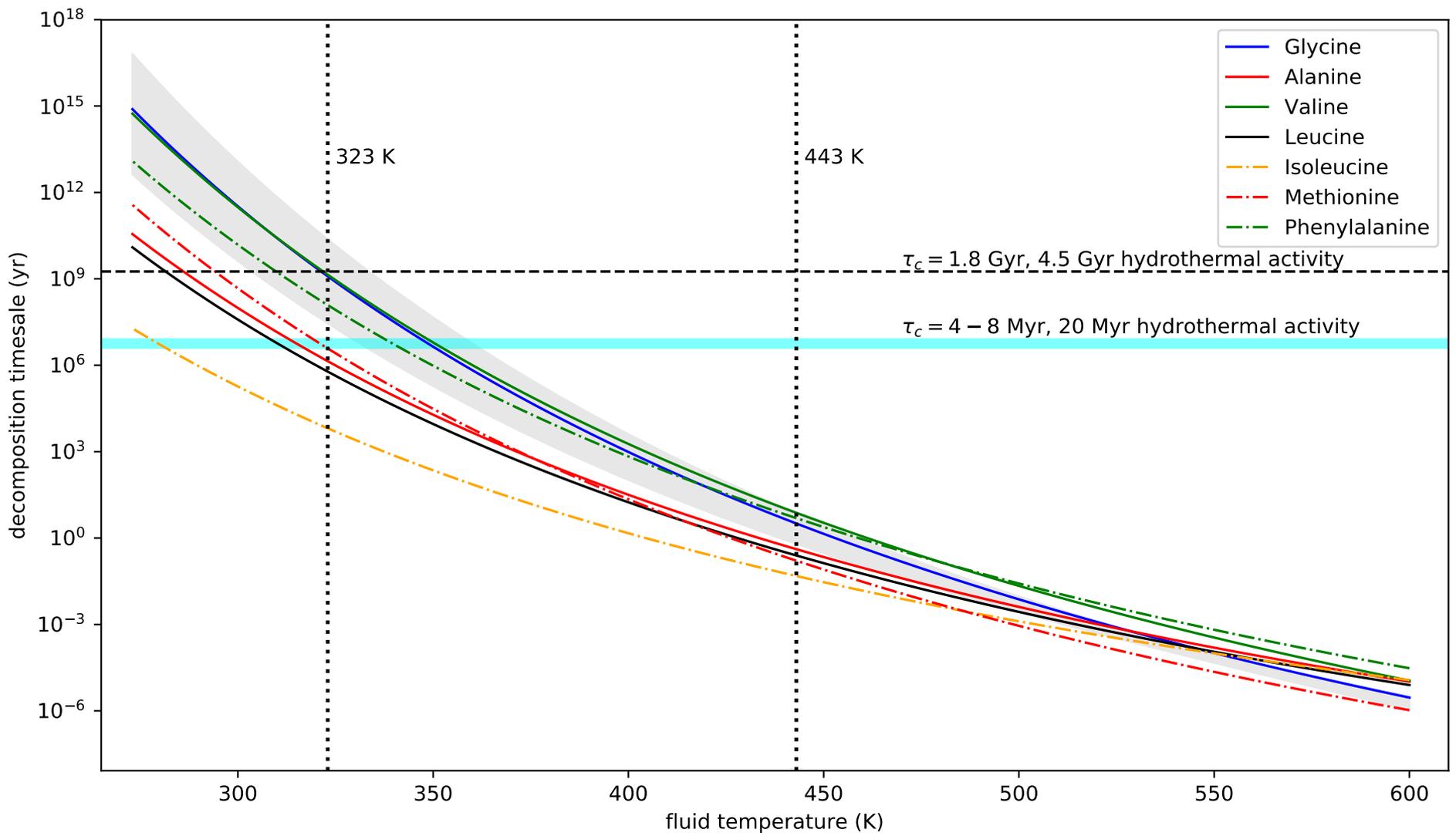
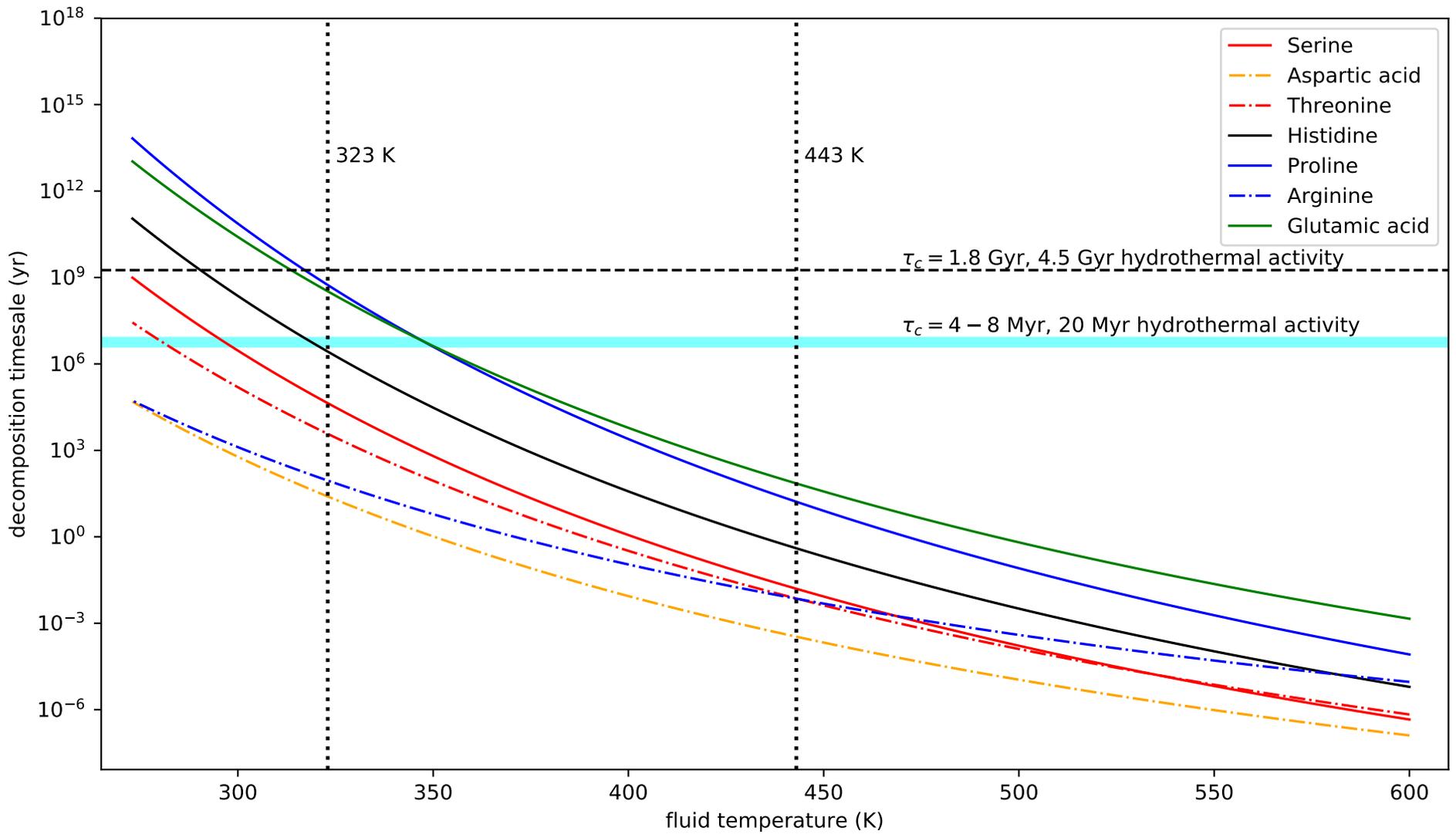

Figure 3